\documentclass[submission,copyright,creativecommons]{eptcs}
\providecommand{\event}{ACL2 2023} 
\usepackage{breakurl}             
\usepackage{hyperref}
\usepackage{underscore}           
\usepackage[nocompress]{cite}

\usepackage{amsmath}
\usepackage{amssymb}
\usepackage{amsbsy}
\usepackage{hhline}
\usepackage{xcolor,colortbl}
\usepackage{mathpartir}
\usepackage{tikz}
\usetikzlibrary{arrows,calc,positioning,backgrounds,fit,shapes,shadows,trees}

\tikzset{
  basic/.style  = {draw,rectangle,font=\sffamily\fontsize{12}{12}\selectfont}, 
  root/.style   = {basic, font=\sffamily\fontsize{12}{12}\selectfont, rounded corners=2pt, thin, align=center,
                   fill=orange!30},
  level 0/.style = {basic, font=\sffamily\fontsize{12}{12}\selectfont, rounded corners=2pt, thin, align=center, fill=blue!10},
  level 1/.style = {basic, font=\sffamily\fontsize{12}{12}\selectfont, thin, align=left, fill=yellow!30},
  level 2/.style = {basic, font=\sffamily\fontsize{12}{12}\selectfont, rounded corners=2pt, thin, align=left,
    fill=yellow!30},
  level 3/.style = {basic, font=\sffamily\fontsize{12}{12}\selectfont, rounded corners=2pt, thin, align=left,
    fill=blue!20, text width=12em},
  level 4/.style = {basic, font=\sffamily\fontsize{12}{12}\selectfont, thin, align=center, 
    fill=green!30},
  level 5/.style = {basic, font=\sffamily\fontsize{12}{12}\selectfont, thin, align=center, 
    fill=orange!30},
  level 6/.style = {basic, rectangle, font=\sffamily\fontsize{12}{12}\selectfont, thin, align=center, 
    fill=cyan!20},
  level 7/.style = {basic, font=\sffamily\fontsize{12}{12}\selectfont, thin, draw=white, align=center},
   boxaround/.style={draw=violet, font=\sffamily\fontsize{12}{12}\selectfont, thick, dashdotted,
     inner sep=0.8em},
 >=latex
}

\usepackage{paralist}
\usepackage{boxedminipage} 
\usepackage{booktabs} 

\newcommand{\eg}{\textit{e.g.}}

\usepackage{scalerel}

\title{Verification of a Rust Implementation of Knuth's Dancing Links using ACL2}

\author{David S. Hardin
\institute{Cedar Rapids, IA USA}
\email{david.s.hardin@gmail.com}}

\def\titlerunning{Verification of Rust Dancing Links using ACL2}
\def\authorrunning{D.S. Hardin}

\begin{document}

\maketitle

\begin{abstract}

``Dancing Links'' connotes an optimization to a circular doubly-linked
list data structure implementation which provides for fast list element
removal and restoration.  The Dancing Links optimization is used primarily
in fast algorithms to find exact covers, and has been popularized by Knuth 
in Volume 4B of his seminal series \emph{The Art of Computer Programming}.
We describe an implementation of the Dancing Links optimization in
the Rust programming language, as well as its formal verification using
the ACL2 theorem prover.  Rust has garnered significant endorsement
in the past few years as a modern, memory-safe successor to C/C++
at companies such as Amazon, Google, and Microsoft, and is being
integrated into both the Linux and Windows operating system kernels.
Our interest in Rust stems from its potential as a hardware/software
co-assurance language, with application to critical systems.  We
have crafted a Rust subset, inspired by Russinoff's Restricted
Algorithmic C (RAC), which we have imaginatively named Restricted
Algorithmic Rust, or RAR.  In previous work, we described our initial
implementation of a RAR toolchain, wherein we simply transpile the
RAR source into RAC.  By so doing,  we leverage a number of existing
hardware/software co-assurance tools with a minimum investment of
time and effort.  In this paper, we describe the RAR Rust subset, describe
our improved prototype RAR toolchain, and detail the design and verification
of a circular doubly-linked list data structure employing the Dancing Links
optimization in RAR, with full proofs of functional correctness accomplished
using the ACL2 theorem prover.

\end{abstract}

\section{Introduction}

The exact cover problem \cite{Knuth4B}, in its simplest form, attempts
to find, for an $n \times m$ matrix with binary elements, all of the subsets of
the rows of the matrix such that all the column sums are exactly one.
This basic notion naturally extends to matrix elements that are in
some numerical range; indeed, the popular puzzle game Sudoku is 
an extended exact cover problem for a $9 \times 9$ matrix with element values 
in the range of 1 to 9, inclusive.

The exact cover problem is NP-complete, but computer scientists have
devised recursive, nondeterministic backtracking algorithms to find exact
covers.  One such procedure is Knuth's Algorithm X, described in
\cite{Knuth4B}.  In this algorithm, elements of the matrix are
connected via circular doubly-linked lists, and individual elements
are removed, or restored, as the algorithm proceeds, undergoing 
backtracking, etc.  As these removals and restorations out of/into the
list are quite common, making these operations efficient is a laudable
goal.  This is where Knuth's ``Dancing Links'' comes in, resulting in
an optimized algorithm for finding exact covers which Knuth calls DLX
(Dancing Links applied to algorithm X).

\section{Dancing Links}

\begin{figure}
\center{
\begin{tikzpicture}[auto, scale=1.2]
\node[level 5] (xold) at (2,3.2) [shape=rectangle,draw,align=center]{Element\\X};
\node[level 5] (yold) at (2,1.7) [shape=rectangle,draw,align=center]{Element\\Y};
\node[level 5] (zold) at (2,0.2) [shape=rectangle,draw,align=center]{Element\\Z};
\node[level 5] (x) at (4.7,3.2) [shape=rectangle,draw,align=center]{Element\\X};
\node[level 1] (y) at (4.7,1.7) [shape=rectangle,draw,align=center]{Element\\Y};
\node[level 5] (z) at (4.7,0.2) [shape=rectangle,draw,align=center]{Element\\Z};
\node[level 5] (xnew) at (7.4,3.2) [shape=rectangle,draw,align=center]{Element\\X};
\node[level 5] (ynew) at (7.4,1.7) [shape=rectangle,draw,align=center]{Element\\Y};
\node[level 5] (znew) at (7.4,0.2) [shape=rectangle,draw,align=center]{Element\\Z};
\draw [->,thick] (xold.230) to (yold.130) node[left, yshift=11pt]{\emph{next}};
\draw [->,thick] (yold.60) to (xold.300) node[right,yshift=-11pt]{\emph{prev}};;
\draw [->,thick] (yold.230) to (zold.130) node[left, yshift=11pt]{\emph{next}};
\draw [->,thick] (zold.60) to (yold.300) node[right,yshift=-11pt]{\emph{prev}};
\draw [->,thick] (x) to [out=210,in=150]  (z);
\draw [->,thick] (z) to [out=30,in=330]  (x);
\draw [shorten >=8 pt, shorten <=0pt,->,thick] (y.56) to (x.307) node[right,yshift=-15pt]{\emph{prev}};;
\draw [shorten >=8 pt, shorten <=0pt,->,thick] (y.231) to (z.132) node[left,yshift=15pt]{\emph{next}};;
\draw [->,thick] (xnew.230) to (ynew.130) node[left, yshift=11pt]{\emph{next}};
\draw [->,thick] (ynew.60) to (xnew.300) node[right,yshift=-11pt]{\emph{prev}};;
\draw [->,thick] (ynew.230) to (znew.130) node[left, yshift=11pt]{\emph{next}};
\draw [->,thick] (znew.60) to (ynew.300) node[right,yshift=-11pt]{\emph{prev}};
\node[level 7] (old) at (2, -1) {(a)};
\node[level 7] (old) at (4.7, -1) {(b)};
\node[level 7] (old) at (7.4, -1) {(c)};
\end{tikzpicture}}
\caption{Dancing Links in action. (a) Portion of a circular
  doubly-linked list prior to a remove operation; (b) After the
  remove operation on element Y; (c) After the restore operation for
  element Y.}
\label{dancing-links-pic}
\end{figure}
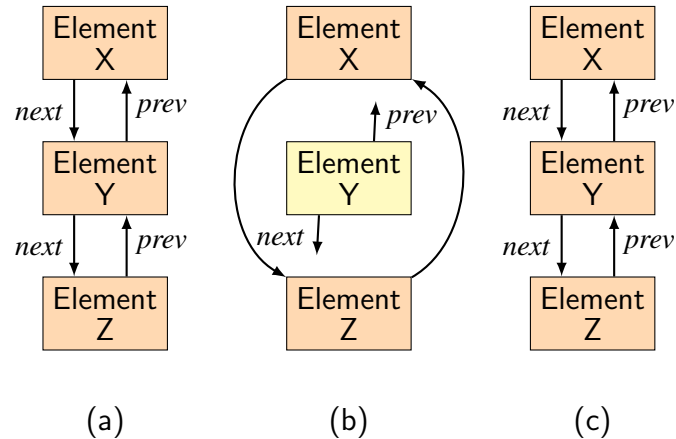

The concept behind Dancing Links is quite simple: when a given element
Y of a list is removed in an exact cover algorithm, it is very likely
that this same element will later be restored.  Thus, rather than
``zero out'' the `previous' and `next' links associated with element
Y, as good programming hygiene would normally dictate, in Dancing
Links, the programmer leaves the link values in place for the removed
element.  The Dancing Links \texttt{remove} operator thus
deletes element Y from the list, setting the 'next' element of the
preceding element X to the following element Z, and setting the
'previous' element of Z to a link to X, but not touching the 'next' and
'previous' links of the removed element Y.  Later on, if Y needs to be
restored, it is simply hooked back in to the list using a simple
\texttt{restore} operator.  In Knuth's words, if one monitors the
list links as the DLX algorithm proceeds, the links appear to `dance', hence the
name.  Knuth's Dancing Links functionality is summarized in
Fig.~\ref{dancing-links-pic}.

\section{The Rust Programming Language}

The Rust programming language has garnered significant interest and
use as a modern, type-safe, memory-safe, and potentially formally
analyzable programming language.  Google \cite{RustAndroid} and
Amazon \cite{SustainableRust} are major Rust adopters, and Linus
Torvalds has commented positively on the near-term ability of the
Rust toolchain to be used in Linux kernel development \cite{RustAndroidArs}.
And after spending decades dealing with a never-ending parade of
security vulnerabilities due to C/C++, which continue to
manifest at a high rate \cite{MSBugs} despite their use of
sophisticated C/C++ analysis tools, Microsoft announced at its
BlueHat 2023 developer conference that is was beginning to rewrite
core Windows libraries in Rust \cite{RustWindows}.

Our interest in Rust stems from its potential as a hardware/software
co-assurance language.  This interest is motivated in part by emerging
application areas, such as autonomous and semi-autonomous
platforms for land, sea, air, and space, that require sophisticated
algorithms and data structures, are subject to stringent
accreditation/certification, and encourage hardware/software co-design
approaches. (For an unmanned aerial vehicle use case illustrating a formal
methods-based systems engineering environment, please consult
\cite{CASEatscale} \cite{case-models-2023}.)  In this paper, we explore the use
of Rust as a High-Level Synthesis (HLS) language \cite{HLS}.

HLS developers specify the high-level abstract behavior of a digital
system in a manner that omits hardware design details such as
clocking; the HLS toolchain is then responsible for ``filling in the details''
to produce a Register Transfer Level (RTL) structure that can be used
to realize the design in hardware. HLS development is thus closer to
software development than traditional hardware design in
Hardware Description Languages (HDLs) such as Verilog or
VHDL.  Most incumbent HLS languages are a subset of C, e.g. Mentor
Graphics' Algorithmic C \cite{AlgoC}, or Vivado HLS by Xilinx
\cite{VivadoHLS}, although other languages have also been used, e.g.
OCaml \cite{Hardcaml}.  A Rust-based
HLS would bring a single modern, type-safe, and memory-safe
expression language for both hardware and software realizations,
with very high assurance.

For formal methods researchers, Rust presents the opportunity 
to reason about application-level logic written in the imperative
style favored by industry, but without the snarls of the 
unrestricted pointers of C/C++.  Much progress has been made to
this end in recent years, to the point that developers can verify the
correctness of common algorithm and data structure code that utilizes
common idioms such as records, loops, modular integers, and the like,
and verified compilers can guarantee that such code is compiled
correctly to binary \cite{cakeml:popl14}.  Particular progress has 
been made in the area of hardware/software co-design algorithms, 
where array-backed data structures are
common \cite{hardin-co-assurance, hardin-rac}.
(NB: This style of programming also addresses one of the shortcomings 
of Rust, namely its lack of support for cyclic data structures.)

As a study of the suitability of Rust as an HLS, we have crafted a
Rust subset, inspired by Russinoff's Restricted Algorithmic C
(RAC) \cite{Russinoff2022}, which we have imaginatively named
Restricted Algorithmic Rust, or RAR \cite{RAR}.  In fact, in our first
implementation of a RAR toolchain, we merely ``transpile'' 
(perform a source-to-source translation of) the
RAR source into RAC.  By so doing, we leverage a number of existing
hardware/software co-assurance tools with a minimum
investment of time and effort.  By transpiling  RAR to RAC, we gain
access to existing HLS compilers (with the help of some simple C
preprocessor directives, we are able to generate code for either the
Algorithmic C or Vivado HLS toolchains).  But most importantly for
our research, we leverage the RAC-to-ACL2 translator that Russinoff
and colleagues at Arm have successfully utilized in
industrial-strength floating point hardware verification.

We have implemented several representative algorithms and data
structures in RAR, including:

\begin{itemize}
\item{a suite of array-backed algebraic data types, previously
    implemented in RAC (as reported in \cite{hardin-rac});}
\item{a significant subset of the Monocypher \cite{monocypher} modern cryptography
    suite, including XChacha20 and Poly1305 (RFC 8439) encryption/decryption,
    BLAKE2b hashing, and X25519 public key cryptography \cite{RAR-ZT}; and}
\item{a DFA-based JSON lexer, coupled with an LL(1) JSON parser.  
   The JSON parser has also been implemented using Greibach
   Normal Form (previously implemented in RAC, as described in
   \cite{formal-filter-synth-langsec}).}
\end{itemize}

The RAR examples created to date are similar to their RAC counterparts
in terms of expressiveness, and we deem the RAR versions somewhat
superior in terms of readability (granted, this is a very subjective evaluation).

In this paper, we will describe the development and formal verification 
of an array-based circular doubly-linked list (CDLL) data structure in
RAR, including the Dancing Links optimization.  Along the way, we will
introduce the RAR subset of Rust, the RAR toolchain, the CDLL example,
and detail our ACL2-based verification techniques, as well as the ACL2
books that we brought to bear on this example.  It is hoped that this
explication will convince the reader of the practicality of RAR as a
high-assurance hardware/software co-design language, as well as the
feasibility of the performing full functional correctness proofs of
RAR code.  We will then conclude with related and future work.

\section{RAC: Hardware/Software Co-Assurance at Scale}
\label{sec:algoc}

In order to begin to realize hardware/software co-assurance
at scale, we have  conducted several experiments employing a
state-of-the-art toolchain, due to Russinoff and O'Leary, and
originally designed for use  in floating-point hardware verification \cite{Russinoff2022}, 
to determine its suitability for the creation of
safety-critical/security-critical applications in various domains.
Note that this toolchain has already demonstrated the capability to
scale to industrial designs in the floating-point hardware design and
verification domain, as it has been used in design verifications for
CPU products at both Intel and Arm.

Algorithmic C \cite{AlgoC} is a High-Level Synthesis (HLS) language,
and is supported by hardware/software co-design environments from
Mentor Graphics, \eg, Catapult \cite{Catapult}.  Algorithmic C defines
C++ header files that enable compilation to both hardware and software
platforms, including support for the peculiar bit widths employed, for
example, in floating-point hardware design.

\begin{figure}
\center{
\begin{tikzpicture}[auto, scale=1.2]
\node[level 5] (Hdr) at (2,3.2) [shape=rectangle,draw,align=center]{Algorithmic C\\Headers};
\node[level 5] (AlgoC) at (2,1.7) [shape=rectangle,draw,align=center]
{Algorithmic  C\\Source};
\node[level 0] (C++) at (1,0) [shape=rectangle,draw,align=center]{C++\\Compiler};
\node[level 0] (HW) at (3,0) [shape=rectangle,draw,align=center]{Hardware\\Synthesis};
\node[level 2] (Xlate) at (4.7,1.7) [shape=rectangle,draw,align=center]{ACL2\\Translator};
\node[level 4] (Lemmas) at (7,3.2) [shape=rectangle,draw,align=center]{Lemmas};
\node[level 2] (ACL2) at (7,1.7) [shape=rectangle,draw,align=center]{ACL2\\Theorem\\Prover};
\node[level 4] (Proofs) at (7,0.1) [shape=rectangle,draw,align=center]{Proofs};
\draw [->,thick] (Hdr) to node {\textit{\#include}} (AlgoC);
\draw [->,thick] (AlgoC) to (C++);
\draw [->,thick] (AlgoC) to (HW);
\draw [->,thick] (AlgoC) to (Xlate);
\draw [->,thick] (Xlate) to (ACL2);
\draw [->,thick] (Lemmas) to (ACL2);
\draw [->,thick] (ACL2) to (Proofs);
\end{tikzpicture}}
\caption{Restricted Algorithmic C (RAC) toolchain.}
\label{RAC:toolchain}
\end{figure}
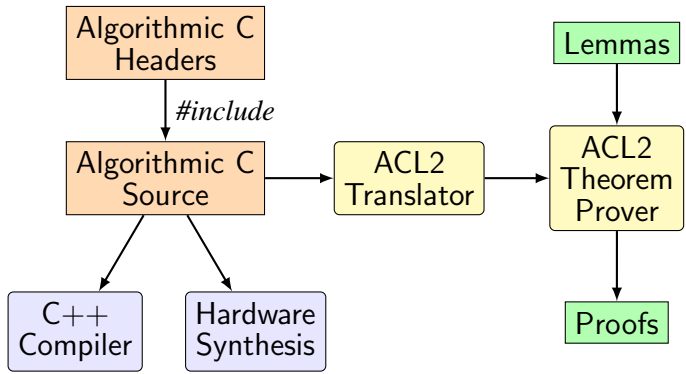

The Russinoff-O'Leary Restricted Algorithmic C (RAC)
toolchain, depicted in Fig.~\ref{RAC:toolchain}, translates a subset of
Algorithmic C source to the Common Lisp subset supported by the 
ACL2 theorem prover, as augmented by Russinoff's Register Transfer
Logic (RTL) books.

\begin{table}
  \center{
  \begin{normalsize}
\begin{tabular}{|c|c|}
  \hline
  \emph{Formal Verification ``Comfort Zone''} & \emph{Real-World Development} \\ \hhline{|=|=|}
  Functional programming & Imperative programming \\ \hline
  Total, terminating functions	& Partial, potentially non-terminating functions \\ \hline
  Non-tail-recursive functions & Loops \\ \hline
  Okasaki-style pure functional algebraic data types & Structs, Arrays \\ \hline
  Infinite-precision Integers, Reals & Modular Integers, IEEE 754 floating point\\ \hline
  Linear Arithmetic & Linear and Non-linear Arithmetic \\ \hline
  Arithmetic or Bit Vectors & Arithmetic \emph{and} Bit Vectors \\ \hline
\end{tabular}
\end{normalsize}
}
\bigskip
\caption{Formal verification vs. real-world development attributes.}
\label{tbl:fm-vs-real-world}
\end{table}

The ACL2 Translator component of Fig.~\ref{RAC:toolchain} provides a
case study in the bridging of Formal Modeling and Real-World
Development concerns, as summarized in
Table~\ref{tbl:fm-vs-real-world}.  The ACL2 translator converts
imperative RAC code to functional ACL2 code.  Loops are translated
into tail-recursive functions, with automatic generation of measure
functions to guarantee admission into the logic of ACL2 (RAC
subsetting rules ensure that loop measures can be automatically
determined).  Structs and arrays are converted into functional ACL2
records.  The combination of modular arithmetic and bit-vector
operations of typical RAC source code is faithfully translated to
functions supported by Russinoff's RTL books.  ACL2 is able to reason
about non-linear arithmetic functions, so the usual concern about
formal reasoning about non-linear arithmetic functions does not apply.  
Finally, the RTL books are quite capable of reasoning about 
a combination of arithmetic and bit-vector operations, which is
a very difficult feat for most automated solvers.

Recently, we have investigated the synthesis of  
Field-Programmable Gate Array (FPGA) hardware directly from
high-level architecture models, in collaboration with
colleagues at Kansas State University.  The goal of this work 
is to enable the generation of high-assurance hardware and/or
software from high-level architectural specifications expressed in the
Architecture Analysis and Design Language (AADL) \cite{feiler-AADL},
with proofs of correctness in ACL2.

\section{Rust and RAR}

The Rust Programming Language \cite{Rust2018} is a modern, high-level
programming language designed to combine the code generation
efficiency of C/C++ with drastically improved type safety and memory
management features.  A distinguishing feature of Rust is that a non-scalar
object may only have one owner.  For example, one cannot assign a
reference to an object in a local variable, and then pass that
reference to a function.  This restriction is similar to those imposed
on ACL2 single-threaded objects (stobjs) \cite{stobj}, with the
additional complexities of enforcing such ``single-owner''
restrictions in the context of a general-purpose, imperative
programming language.  The Rust runtime performs array
bounds checking, as well as arithmetic overflow checking (the latter
can be disabled by a build environment setting).

In most other ways, Rust is a fairly conventional modern programming
language, with interfaces (called traits), lambdas (termed closures),
and pattern matching, as well as a macro capability.  Also in keeping
with other modern programming language ecosystems, Rust features a
language-specific build and package management sytem, named \texttt{cargo}.

\subsection{Restricted Algorithmic Rust}

As we wish to utilize the RAC toolchain as a backend in our initial
work, Restricted Algorithmic Rust is semantically equivalent to
RAC.  Thus, we adopt the same semantic restrictions as described in
Russinoff's book.  Additionally, in order to enable translation to RAC,
as well as to ease the transition from C/C++, RAR supports a
commonly used macro that provides a C-like \emph{for} loop in Rust.  
Note that, despite the restrictions, RAR code is proper Rust; it
compiles to binary using the standard Rust compiler.

\begin{figure}
\center{
\begin{tikzpicture}[auto, scale=1.2]
\node[level 6] (Rust) at (-3.2,1.7) [shape=rectangle,draw,align=center]{Rust\\Source};
\node[level 2] (Plexi) at (-0.8,1.7) [shape=rectangle,draw,align=center]{Plexi\\Transpiler};
\node[level 5] (Hdr) at (2,3.2) [shape=rectangle,draw,align=center]{Algorithmic C\\Headers};
\node[level 5] (AlgoC) at (2,1.7) [shape=rectangle,draw,align=center]{Algorithmic  C\\Source};
\node[level 0] (C++) at (1,0) [shape=rectangle,draw,align=center]{C++\\Compiler};
\node[level 0] (HW) at (3,0) [shape=rectangle,draw,align=center]{Hardware\\Synthesis};
\node[level 2] (Xlate) at (4.7,1.7) [shape=rectangle,draw,align=center]{ACL2\\Translator};
\node[level 4] (Lemmas) at (7,3.2) [shape=rectangle,draw,align=center]{Lemmas};
\node[level 2] (ACL2) at (7,1.7) [shape=rectangle,draw,align=center]{ACL2\\Theorem\\Prover};
\node[level 4] (Proofs) at (7,0.1) [shape=rectangle,draw,align=center]{Proofs};
\draw [->,thick] (Rust) to (Plexi);
\draw [->,thick] (Plexi) to (AlgoC);
\draw [->,thick] (Hdr) to node {\textit{\#include}} (AlgoC);
\draw [->,thick] (AlgoC) to (C++);
\draw [->,thick] (AlgoC) to (HW);
\draw [->,thick] (AlgoC) to (Xlate);
\draw [->,thick] (Xlate) to (ACL2);
\draw [->,thick] (Lemmas) to (ACL2);
\draw [->,thick] (ACL2) to (Proofs);
\end{tikzpicture}}
\caption{Restricted Algorithmic Rust (RAR) toolchain.}
\label{RAR:toolchain}
\end{figure}
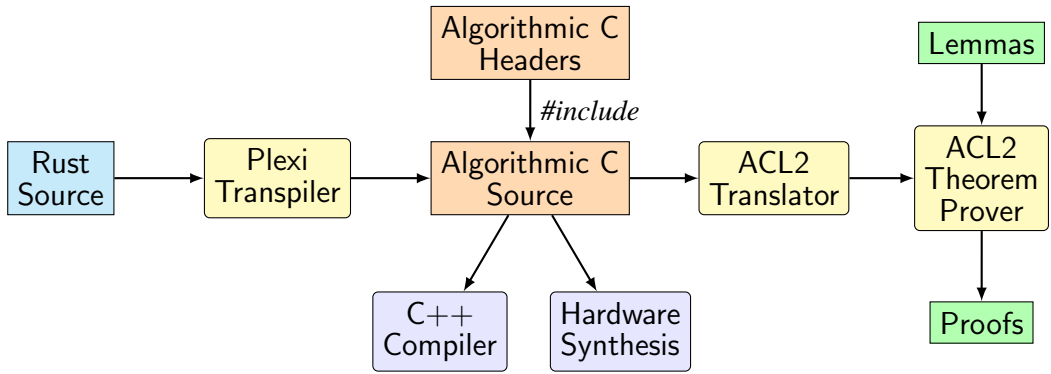

RAR is transpiled to RAC via a source-to-source translator, as
depicted in Fig.~\ref{RAR:toolchain}.  Our transpiler is based on
the \texttt{plex} parser and lexer generator \cite{plex} source code.
We thus call our transpiler \emph{Plexi}, a nickname given to a famous
(and now highly sought-after) line of Marshall guitar amplifiers of
the mid-1960s.  Plexi performs lexical and syntactic transformations
that convert RAR code to RAC code.  Recent improvements in the
\texttt{plexi} tool include better handling of array declarations, as
well as providing support for Rust \texttt{const} declarations.

The generated RAC code can then be compiled using a C/C++ compiler,
fed to an HLS-based FPGA compiler, as well as translated to ACL2 via
the RAC ACL2 translator, as illustrated in Fig.~\ref{RAR:toolchain}.

\section{Dancing Links in Rust}

In this section, we describe an array-based circular doubly-linked
list (CDLL) employing Knuth's ``Dancing Links'' optimization, realized
using our RAR Rust subset.  The CDLL data structure implementation
constitutes over 700 lines of Rust code, which becomes 890 lines of
code when translated to ACL2.

\subsection{Definitions}
\label{CDLLDef}

First, we present the basic RAR declaration for the CDLL.

\begin{verbatim}

const CDLL_MAX_NODE1: usize = 8191;
const CDLL_MAX_NODE: usize = CDLL_MAX_NODE1 - 1;

#[derive(Copy, Clone)]
struct CDLLNode {
  alloc: u2,
  val: i64,
  prev: usize,
  next: usize,
}

#[derive(Copy, Clone)]
struct CDLL {
  nodeHd: usize,
  nodeCount: usize,
  nodeArr: [CDLLNode; CDLL_MAX_NODE1],
}
\end{verbatim}

Rust data structure declarations are similar to those in C, but struct
elements are declared by specifying the element name, followed by
the \texttt{:} separator, then the element type.  Also note that Rust
pragmas may be given using  the \texttt{derive} attribute.  In the
declaration above, the array \texttt{nodeArr} holds the list element
nodes.  Each element has \texttt{next} and \texttt{prev} indices.
Note that indices in Rust are normally declared to be of the
\texttt{usize} type.  Note also that by using array indices instead of
references, we get around Rust ownership model issues with circular
data structures.  The \texttt{alloc} field of the \texttt{CDLLNode}
structure is declared to be a two bit unsigned field, but its only
allowed values are two non-zero values: 2 (not currently allocated),
and 3 (allocated).  The reason for this has to do with the details of
ACL2 untyped record reasoning, which will be discussed in
Section~\ref{sec:acl2-trans}.

The Dancing Links operators \texttt{cdll_remove} and
\texttt{cdll_restore} are presented in Figures \ref{cdll-remove}
and \ref{cdll-restore}, respectively.  Rust functions begin with
the \texttt{fn} keyword, followed by the function name, a
parenthesized list of parameters, the \texttt{->} (returns) symbol,
the return type name, followed by the function body (delimited by
a curly brace pair).  A function parameter list element consists of
the parameter name, the \texttt{:} symbol, then the parameter type.
Additional parameter modifiers, for example \texttt{mut}, may be present
to indicate that the parameter is changed in the function body.
Within the function body, the syntax is similar to other C-like
languages, but local variable declarations begin with
\texttt{let}, and use the variable name, \texttt{:}, variable type
declaration syntax.  A local variable declaration may also require the
\texttt{mut} modifier if that local variable is updated after its initialization.

\begin{figure*}
\begin{verbatim}
fn CDLL_remove(n: usize, mut CDObj: CDLL) -> CDLL {
  if (n > CDLL_MAX_NODE) {
    return CDObj;
  } else {
    if (n == CDObj.nodeHd) {  // Can't remove head
      return CDObj;
    } else {
      if (CDObj.nodeCount < 3) {  // Need three elements for remove to work
        return CDObj;
      } else {
        let nextNode: usize = CDObj.nodeArr[n].next;
        let prevNode: usize = CDObj.nodeArr[n].prev;

        CDObj.nodeArr[prevNode].next = nextNode;
        CDObj.nodeArr[nextNode].prev = prevNode;

        CDObj.nodeCount = CDObj.nodeCount - 1;

        return CDObj;
      }
    }
  }
}
\end{verbatim}
\hrulefill
\caption{\texttt{cdll_remove()} function in RAR.}
\label{cdll-remove}
\end{figure*}

\begin{figure*}
\begin{verbatim}
fn CDLL_restore(n: usize, mut CDObj: CDLL) -> CDLL {
  if (n > CDLL_MAX_NODE) {
    return CDObj;
  } else {
    if (n == CDObj.nodeHd) {  // Can't restore head
      return CDObj;
    } else {
      if ((CDObj.nodeCount < 2) ||        // Need two elements for restore to work
          (CDObj.nodeCount == CDLL_MAX_NODE1))  {  // Can't restore to a full list
        return CDObj;
      } else {

        let prevNode: usize = CDObj.nodeArr[n].prev;
        let nextNode: usize = CDObj.nodeArr[n].next;

        CDObj.nodeArr[prevNode].next = n;
        CDObj.nodeArr[nextNode].prev = n;

        CDObj.nodeCount = CDObj.nodeCount + 1;

        return CDObj;
      }
    }
  }
}
\end{verbatim}
\hrulefill
\caption{\texttt{cdll_restore()} function in RAR.}
\label{cdll-restore}
\end{figure*}

\subsection{Translation to ACL2}
\label{sec:acl2-trans}

We use \texttt{Plexi} to transpile the RAR source to RAC (not
shown), then use the RAC translator to convert the resulting RAC
source to ACL2.  The translation of \texttt{cdll_restore()} appears in
Fig.~\ref{cdll-restore-acl2}.

\begin{figure*}
\begin{verbatim}
(DEFUND CDLL_RESTORE (N CDOBJ)
   (IF1 (LOG> N (CDLL_MAX_NODE))
        CDOBJ
        (IF1 (LOG= N (AG 'NODEHD CDOBJ))
             CDOBJ
             (IF1 (LOGIOR1 (LOG< (AG 'NODECOUNT CDOBJ) 2)
                           (LOG= (AG 'NODECOUNT CDOBJ)
                                 (CDLL_MAX_NODE1)))
                  CDOBJ
                  (LET* ((PREVNODE (AG 'PREV (AG N (AG 'NODEARR CDOBJ))))
                         (NEXTNODE (AG 'NEXT (AG N (AG 'NODEARR CDOBJ))))
                         (CDOBJ (AS 'NODEARR
                                    (AS PREVNODE
                                        (AS 'NEXT
                                            N (AG PREVNODE (AG 'NODEARR CDOBJ)))
                                        (AG 'NODEARR CDOBJ))
                                    CDOBJ))
                         (CDOBJ (AS 'NODEARR
                                    (AS NEXTNODE
                                        (AS 'PREV
                                            N (AG NEXTNODE (AG 'NODEARR CDOBJ)))
                                        (AG 'NODEARR CDOBJ))
                                    CDOBJ)))
                        (AS 'NODECOUNT
                            (+ (AG 'NODECOUNT CDOBJ) 1)
                            CDOBJ))))))
\end{verbatim}
\hrulefill
\caption{\texttt{cdll_restore()} function translated to ACL2 using the RAC tools.}
\label{cdll-restore-acl2}
\end{figure*}

The first thing to note about Fig.~\ref{cdll-restore-acl2} is that,
even though we are two translation steps away from the original
RAR source, the translated function is nonetheless quite readable,
which is a rare thing for machine-generated code.  Another notable
observation is that struct and array `get' and `set' operations become
untyped record operators, \texttt{AG} and \texttt{AS}, respectively ---
these are slight RAC-specific customizations of the usual ACL2 untyped
record operators. Further, \texttt{IF1} is a RAC-specific  macro, and \texttt{LOG>}, \texttt{LOG=},
\texttt{LOG<}, and \texttt{LOGIOR1}  are all RTL functions.  Thus, much of the proof effort
involved with RAR code is reasoning about untyped records and RTL --- 
although not a lot of RTL-specific knowledge is needed, at least in our
experience.

One aspect of untyped records that can be tricky is that record
elements that take on the default value are not explicitly stored
in the association list for the record.  For RAC untyped records, that
default value is zero.  Thus, it is easy for a given record to attain a
\texttt{nil} value.  When reasoning about arrays of such records,
it is often desirable to be able to state that the array size remains
constant.  Thus, for example, for the \texttt{CDLL} array 
\texttt{nodeArr} of Section~\ref{CDLLDef}, we ensure that all
\texttt{CDLLNode} elements of that array are non-nil by making sure that
the \texttt{alloc} fields of the \texttt{CDLLNode} elements are always
non-zero (2 or 3).

\subsection{Dancing Links Theorems}\label{CDLLThms}

Once we have translated the circular doubly-linked list functions into
ACL2, we can begin to prove theorems about the data structure
implementation.  We begin by defining  a ``well-formedness''
predicate for CDLLs.

\begin{verbatim}
(defun cdllnodeArrp-helper (arr j)
  (cond ((not (true-listp arr)) nil)
        ((null arr) t)
        ((not (and (integerp j) (<= 0 j))) nil)
        ((not (consp (car arr))) nil)
        ((not (= (car (car arr)) j)) nil)
        ((not (cdllnodep (cdr (car arr)))) nil)
        (t (cdllnodeArrp-helper (cdr arr) (1+ j)))))

(defun cdllnodeArrp (arr)
  (cdllnodeArrp-helper arr 0))

(defun cdllp (Obj)
  (and (integerp (ag 'nodeHd Obj))
       (<= 0 (ag 'nodeHd Obj))
       (<= (ag 'nodeHd Obj) (CDLL_MAX_NODE))
       (integerp (ag 'nodeCount Obj))
       (<= 0 (ag 'nodeCount Obj))
       (<= (ag 'nodeCount Obj) (CDLL_MAX_NODE1))
       (cdllnodeArrp (ag 'nodeArr Obj))
       (= (len (ag 'nodeArr Obj)) (CDLL_MAX_NODE1))))
\end{verbatim}

Given this definition of a good CDLL state, we can prove
functional correctness theorems for  Dancing Links operations,
of the sort stated below.  Note that this proof requires some
detailed well-formedness hypotheses related to the \texttt{prev} and
\texttt{next} indices for the nth element:

\begin{verbatim}
(defthm restore-of-remove--thm
 (implies
  (and (cdllp Obj)
       (good-nodep n Obj)
       (not (= n (ag 'nodeHd Obj)))
       (>= (ag 'nodeCount Obj) 3))
  (= (CDLL_restore n (CDLL_remove n Obj))
     Obj)))
\end{verbatim}

ACL2 performs the correctness proof for this \texttt{cdll_restore} of
\texttt{cdll_remove} theorem automatically.  In addition to the
Dancing Links operator proofs, we have proved approximately 160
theorems related to the CDLL data structure, including theorems about
\texttt{cdll_cns()} (\texttt{cons} equivalent), \texttt{cdll_rst()}
(\texttt{cdr} equivalent), \texttt{cdll_snc()} (add to end of data structure),
\texttt{cdll_tsr()} (delete from end of data structure),
\texttt{cdll_nth()}, etc.  All of these proofs will be made publicly
available in the ACL2 workshop books repository.

\section{Related Work}\label{comparison}

A number of domain-specific languages targeting both hardware and
software realization, and providing support for formal verification,
have been created.  Cryptol \cite{Browning10:chapter}, for example,
has been employed  as a ``golden spec'' for the evaluation of cryptographic
implementations, in which automated tools perform equivalence
checking between the Cryptol spec for a given algorithm, and the
VHDL implementation.

Formal verification systems for Rust include Creusot \cite{creusot},
based on WhyML; Prusti \cite{prusti}, based on the Viper verification
toolchain; and RustHorn \cite{rusthorn}, based on constrained
Horn clauses.  AWS is developing a model-checker for Rust,
Kani \cite{kani}.  Additionally, Carnegie-Mellon University is
developing Verus, an SMT-based tool for formally 
verifying Rust programs \cite{CMUVerus}.  With Verus, programmers
express proofs and specifications using Rust syntax, allowing proofs 
to take advantage of Rust's linear types and borrow checking.  It will
be interesting to attempt the sorts of correctness proofs achievable
on our system using these verification tools.

\section{Conclusion}\label{Conclusion}

We have developed a prototype toolchain to allow the Rust
programming language to be used as a hardware/software co-design
and co-assurance language for critical systems, standing on the shoulders of 
Russinoff's team at Arm, and all the great work they have
done on Restricted Algorithmic C.  We have demonstrated the ability to
establish the correctness of several practical data structures commonly found
in high-assurance systems (\eg, array-backed singly-linked lists,
doubly-linked lists, stacks, and dequeues) through automated formal verification,
enabled by automated source-to-source translation from Rust to RAC to
ACL2, and have detailed the specification and verification of one such
data structure, a circular doubly-linked list employing Knuth's
``Dancing Links'' optimization.  We have also successfully
applied our toolchain to cryptography and data format filtering
examples typical of the sorts of algorithms that one encounters in
critical systems development.

In future work, we will continue to develop our toolchain, increasing
the number of Rust features that we can support in the RAR subset, as well
as continuing to improve the ACL2 verification libraries in order to
increase the ability to discharge RAR correctness proofs
automatically.  We will also continue to work with our colleagues
at Kansas State University on the direct synthesis and verification of
RAR code from architectural models, as well as working with colleagues
at the University of Kansas on verified synthesis of Rust code from
high-level attestation protocol specifications written using the
Coq theorem prover.

\section{Acknowledgments}\label{Acknowledgments}

Many thanks to Donald Knuth for his detailed study of exact cover
problems in general, and the ``Dancing Links'' optimization in
particular, that can now be found in Volume 4B of his seminal series,
\emph{The Art of Computer Programming}.  It was a pleasure discovering
this particular corner of Computer Science, beginning when the author
accidentally stumbled upon a previously recorded Knuth
``Christmas Lecture'' on the subject in late 2022.

Previous foundational work on hardware/software co-assurance in Rust
was funded by DARPA contract HR00111890001. The views, opinions and/or findings
expressed are those of the authors and should not be interpreted as
representing the official views or policies of the Department of
Defense or the U.S. Government.

Many thanks to David Russinoff of Arm for developing and improving the
RAC toolchain, without which most of the current work would not be
possible.  Thanks also go to the anonymous reviewers for their 
insightful comments.

\bibliographystyle{eptcs} 
\bibliography{biblio}

\end{document}